\documentclass[fleqn,10pt]{wlscirep}
\title{Visualising Berry phase and diabolical points in a quantum exciton-polariton billiard}

\author[1]{E. Estrecho}
\author[1]{T. Gao}
\author[2]{S. Brodbeck}
\author[2]{M. Kamp}
\author[2]{C. Schneider}
\author[2,3]{S. H\"ofling}
\author[1]{A. G. Truscott}
\author[1,*]{E. A. Ostrovskaya}

\affil[1]{Research School of Physics and Engineering, The Australian National University, Canberra, ACT 2601, Australia}
\affil[2]{Technische Physik and Wilhelm-Conrad-R\"ontgen Research Center for Complex Material Systems, Universit\"at W\"urzburg, Am Hubland, D-97074 W\"urzburg, Germany}
\affil[3]{SUPA, School of Physics and Astronomy, University of St Andrews, St Andrews KY16 9SS, UK}

\affil[*]{elena.ostrovskaya@anu.edu.au}

\begin{abstract}
Diabolical points (degeneracies) can naturally occur in spectra of two-dimensional quantum systems and classical wave resonators due to simple symmetries. Geometric Berry phase is associated with these spectral degeneracies. Here, we demonstrate a diabolical point and the corresponding Berry phase in the spectrum of hybrid light-matter quasiparticles -- exciton-polaritons in semiconductor microcavities. It is well known that sufficiently strong optical pumping can drive exciton-polaritons to quantum degeneracy, whereby they form a macroscopically populated quantum coherent state similar to a Bose-Einstein condensate. By pumping a microcavity with a spatially structured light, we create a two-dimensional quantum billiard for the exciton-polariton condensate and demonstrate a diabolical point in the spectrum of the billiard eigenstates. The fully reconfigurable geometry of the potential walls controlled by the optical pump enables a striking experimental visualisation of the Berry phase associated with the diabolical point. The Berry phase is observed and measured by direct imaging of the macroscopic exciton-polariton wavefunctions.
\end{abstract}
\begin{document}

\flushbottom
\maketitle

\thispagestyle{empty}

\section*{Introduction}

Geometric Pancharatnam-Berry or Berry phase in wave systems is a phenomenon of accumulation of phase of the energy eigenstates during cyclic adiabatic evolution\cite{Berry_phase_1,Berry_phase_2}. In Hermitian quantum systems, the geometric connection of the states of a continuously evolving system with time-varying parameters coincides with the connection of stationary eigenstates in the space of the system's parameters \cite{Arnold}. The Berry phase is therefore a direct consequence of topological properties of the system's eigenvalue surfaces in the parameter space. Since its discovery, Berry phase has been shown to arise in a wide range of physical systems, where it is associated with measurable physical effects. The notable examples are molecular dynamics \cite{Yarkony1996}, polarization rotation and optical spin-Hall effect \cite{Berry_phase_optics,HasmanNP2008,Hasman2008}, as well as orbital magnetism and Hall effects of electronic states in solid-state systems \cite{Berry_phase_solid,Berry_phase_rashba,Berry_phase_graphene}. Moreover, it was speculated that geometric phases in quantum systems could be useful for information processing because of intrinsic topological protection \cite{Moon2009,MoonAPS}.

The most simple and intuitive form of geometric phase in a quantum wave system is associated with spectral degeneracies, i.e. conical intersections of energy surfaces of two eigenstates. Such diabolical points in the space of parameters are most straightforwardly engineered in quantum billiards -- hard-wall resonators for quantum waves -- with spectral degeneracies \cite{Berry1984,Korsch1983}. In the simplest case of a two-level degeneracy, the eigenstates of the billiard accumulate a geometric Berry phase of $\pi$ when a diabolical point is encircled in the space of two parameters responsible for the system's shape. The geometric phase accumulation manifests itself in the rotation of nodal lines of the corresponding wavefunctions \cite{Berry1984,Korsch1983}. This geometric phase has been observed in macroscopic classical microwave resonators \cite{Lauber1994}, and nanoscale electron quantum corrals \cite{Moon2009,MoonAPS}. 

In this work, we visualise the geometric phase of energy eigenstates of a macroscopic quantum system composed of microcavity exciton-polaritons, which bridges the gap between macroscopic classical and nanoscopic quantum waves. Exciton-polaritons are bosonic quasiparticles arising via hybridisation of strongly coupled excitons photons in a semiconductor microcavity \cite{Microcavity_book}. Driven by an optical pump, they undergo bosonic condensation into a macroscopically occupied quantum coherent state \cite{Deng_02,BEC06,BEC07,Deng_10,CiutiREV13,YamamotoREV14}, whose energy and wavefunction can be directly imaged via microcavity photoluminescence. Thus, exciton-polariton condensate is a macroscopic quantum system that spans a range of regimes between classical and quantum waves, while lending itself to optical manipulation and observation {\em in-situ}.

In order to visualise the Berry phase, we construct a two-level system with a diabolical point in the spectrum by confining condensed exciton-polaritons in a quantum billiard  -- an optically-defined resonator for macroscopic quantum waves \cite{Gao2015}. The walls of the resonator are induced by an spatially structured optical pump with the energy tuned far above the energy of the condensing exciton-polaritons. The pump therefore creates a spatially localised reservoir of high-energy excitonic quasiparticles that feeds the exciton-polariton condensate by stimulated scattering \cite{BEC06}. The reservoir interacts with exciton-polaritons repulsively, thus inducing a potential wall \cite{Bloch2010,Tosi2012}. Above the critical pumping enabling the exciton-polariton condensation, the condensate occupies multiple energy eigenstates of the billiard. The shape and deformations of the walls enclosing the billiard are precisely controlled by an optical pump \cite{Gao2015}, thus enabling us to choose two neighbouring energy states and drive them to degeneracy in the form of a diabolical point. Moreover, this precise control over the billiard shape allows us to encircle the diabolical point along a closed path in a plane of two deformation parameters.

In analogy with nanoscopic experiments on surface electrons in quantum corrals \cite{Moon2009}, the Berry phase in our experiments is observed by direct imaging of the nodal patterns of the exciton-polariton wavefunctions, without the need to perform interferometry. 

\section*{Results}

Classical billiards are two-dimensional areas of arbitrary shape surrounded by hard walls of infinite height, where a classical particle exhibits ballistic motion within elastically reflecting boundaries \cite{Robnik}. Quantum counterparts of classical billiards have become a paradigmatic system for studies of transition from integrable (regular) to chaotic behaviour in quantum regime \cite{Berry_sinai}. Square billiards are completely integrable and exhibit regular classical motion. Consequently, systematic (arising from symmetry) degenerate states are abundant in square quantum billiards \cite{Korsch1983,Berry_sinai}, as given by the energy eigenvalues $E_{n,m}=E_0 (n^2+m^2 )/2$, where $E_0$ is the ground state energy, and $n,m=1,2,3\dots$ are quantum numbers. For every combination of $n$ and $m$, there is a two-fold degeneracy of the eigenstates $|n,m\rangle$. Any deformation of the square breaks this symmetry and lifts the degeneracy thus splitting the energy of the previously degenerate states. Conversely, by using two deformation parameters, accidental degeneracies of energy levels can be found without restoring any reflection or rotational symmetry of the system. 

\begin{figure}[ht]
\centering
\includegraphics[width=17cm]{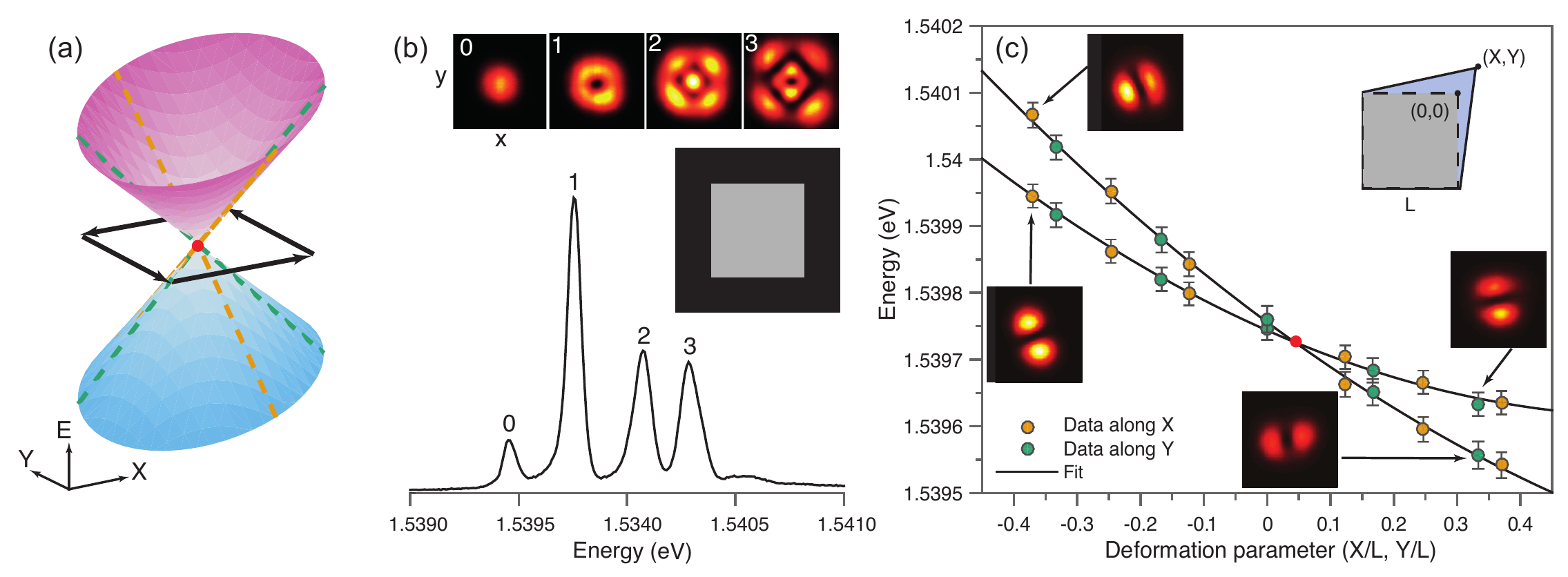}
\caption{(a) The theoretical energy surfaces of the first two excited states denoted by $|1\rangle$ and $|2\rangle$ of an infinite-well, hard-wall square billiard in the $X-Y$ parameter space showing the loop (black arrows) used to obtain the Berry phase. The energy scales with the area of the square and is normalised with respect to the ground state. (b) Typical experimental spectra of exciton-polaritons in a square billiard (photoluminescence intensity, a.u.). The numbered peaks correspond to: $0$ -- a ground state,  $1$ -- the two lowest-order excited states $|1\rangle$ and $|2\rangle$ at the point of degeneracy, $2$ and $3$ -- higher-order states. Inset: (top) The corresponding real space images of the ground and excited states, and (bottom) the spatial shape of the square billiard. Black area corresponds to the optical pump shaped by a DMD (see Methods). (c) The experimentally measured energy corresponding to the dashed lines on the surfaces in (a). Positions of the individual energy levels are determined from the spectroscopic peaks shown in (b). The intersection (red dot) is slightly off the origin since the billiard in the experiment was not perfectly square. The four images shown are the measured real space $(x,y)$ probability densities of the two eigenstates away from the diabolical point. Inset: The deformation parameters $(X,Y)$ shown as coordinates of the deformed corner. Negative values correspond to the shift inwards with respect to the square boundary indicated by the dashed contour; this boundary corresponds to the inner area of the billiard shown in the inset panel of (b).}
\label{Fig1}
\end{figure}

The energy surfaces in parameter space of degenerate states can form a conical intersection -- a diabolical point,  as demonstrated in Figure \ref{Fig1}(a). The energy surfaces shown in Figure \ref{Fig1}(a) are calculated by finding energy eigenvalues in a slightly non-square billiard (one side longer by $10\%$) bordered by hard, infinite walls.  This slightly asymmetric billiard displays no degeneracy for several lowest-lying energy eigenstates. One corner of the billiard is then shifted, as shown in the inset of Figure \ref{Fig1}(c), and its coordinates $X$, $Y$ are the two deformation parameters used to find an accidental degeneracy.

In the vicinity of a diabolical point the behaviour of the energy surfaces can be modelled by a $2\times2$ Hamiltonian of a generic two-level system given by:
\begin{equation}
 H=
\begin{pmatrix} 
\cos \phi    &   \sin \phi \\
\sin \phi      &  -\cos \phi
\end{pmatrix}
\label{eq1}
\end{equation}
where $X=r\cos\phi$ and $Y=r\sin\phi$, and $\phi$ is the deformation angle. The eigenvalues of this Hamiltonian, $E_{\pm}=\pm r$, form two intersecting circular cones in the $E-X-Y$ space similar to that shown in Figure \ref{Fig1}(a). The eigenvectors are linear combinations of the degenerate eigenstates and are functions of the angle $\phi$ only:
\begin{equation}
|+(\phi)\rangle=\begin{pmatrix} \cos(\phi/2) \\ \sin(\phi/2) \end{pmatrix} \qquad |-(\phi)\rangle=\begin{pmatrix} -\sin(\phi/2) \\ \cos(\phi/2) \end{pmatrix}
\label{eq2}
\end{equation}    
It is easy to show that, if we follow a loop that encloses a diabolical point in the parameter plane, a $\pi$-phase shift, i.e. the Berry phase, will result, i.e. $|\pm(\phi+2\pi)\rangle=\exp(i\pi) |\pm(\phi)\rangle$. This means that it takes two full rotations around the point of degeneracy to restore the original wavefunction. 

In our experiment, we create the quantum billiard for exciton-polariton waves by employing the concept of optically-defined polariton potentials \cite{Bloch2010,Tosi2012,Baumberg2013,Lagoudakis2013,Dall2014,Lagoudakis2015,Gao2015,Snoke2016,Review2016} (see Methods for details). These potentials are induced in the quantum well embedded in a microcavity due to the strong Coulomb repulsion between the high-energy excitonic reservoir created by an optical pump and Bose-condensed exciton-polaritons. The spatial distribution of the reservoir particles is defined by the intensity distribution of the optical pump, which, in turn, can be shaped by a spatial light modulator\cite{Tosi2012,Baumberg2013,Lagoudakis2013,Dall2014,Gao2015}, a digital micromirror device (DMD) being the instrument of choice in our experiment. Above a threshold pumping intensity, exciton-polaritons condense into a macroscopic phase-coherent state. The energy spectra and spatial probability distribution for the macroscopic wavefunctions of the exciton-polariton condensate are directly imaged and analysed via the microcavity photoluminescence, as described in Methods. 

For the pump powers above condensation threshold, we find that the exciton-polariton condensate occupies several energy states which are represented by well separated lines in the spectrum, as seen in Figure \ref{Fig1}(b). Due to a slight asymmetry in the pumping profile, we found that the first and second excited states which we denote $|1\rangle$ and $|2\rangle$ are not degenerate at $(0,0)$, i.e. when the corner is not shifted, see Figure \ref{Fig1}(c). These states correspond to the degenerate states $|1,2\rangle$ and $|2,1\rangle$ in an ideal square billiard. By shifting the position of the corner, we can drive the two energy levels to become degenerate at a particular point in $X-Y$ parameter plane, which is a diabolical point. For the billiard we used, the degeneracy occurs at $(0.05,0.05)$ as deduced from the experimental fit of the energy curves in Figure \ref{Fig1}(c). To confirm that there is indeed a true degeneracy, a Berry phase of $\pi$ should be observed as a result of a cyclic change of the parameters along a closed planar contour enclosing the diabolical point. 

\begin{figure}[ht]
\centering
\includegraphics[width=16cm]{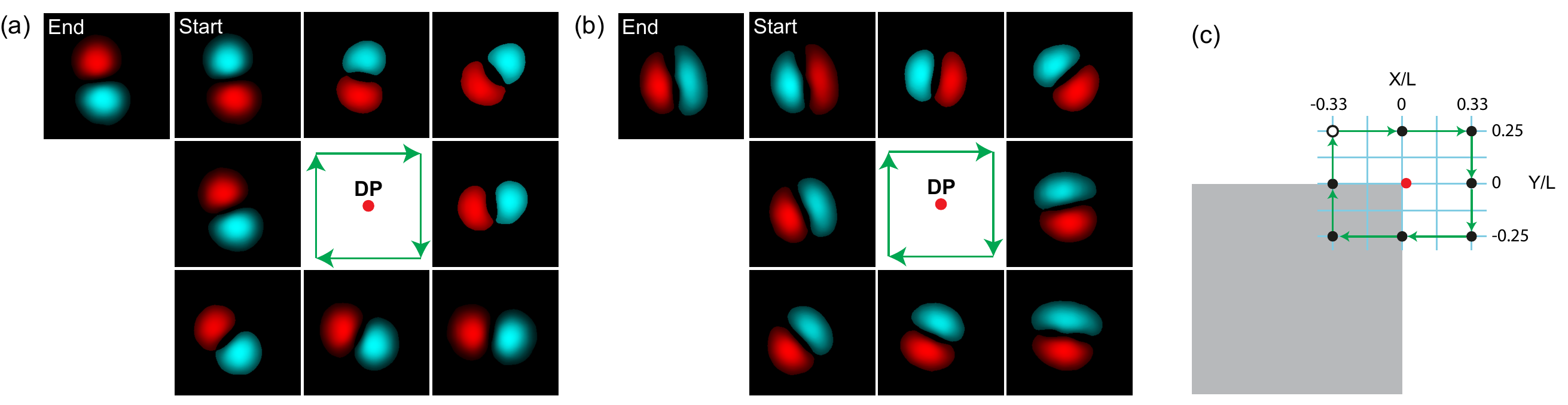}
\caption{Encircling the diabolical point. Panels (a,b) show the change in the non-dynamical phase of the first (a) and second (b) eigenstates of the polariton billiard. An initial phase was assigned to each lobe of the wavefunction at the start; colour contrast corresponds to the $\pi/2$ phase difference. The phase of each successive probability density is set in such a way that it exhibits a smooth transition from the preceding point. After one loop, the phase of each state is flipped, and a Berry phase of $\pi$ is accumulated. (c) Sequence of the billiard deformations $(X,Y)$ corresponding to the loops around the diabolical point, as shown in (a,b). The red dot marks the point of accidental degeneracy, and the open dot marks the start and the end point of the respective loops.}
\label{Fig2}
\end{figure}

\begin{figure}[ht]
\centering
\includegraphics[width=17cm]{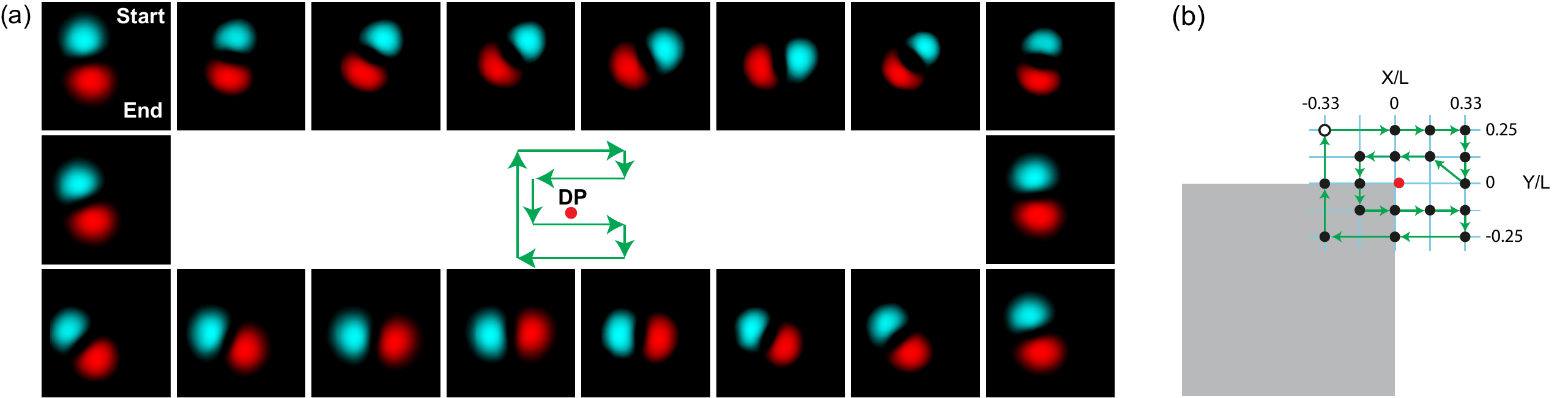}
\caption{Avoiding the diabolical point. (a) The evolution of the phase of the first excited eigenstate in a loop without enclosing the diabolical point in the parameter plane. After a complete loop, there is no phase shift. Inset shows schematics of the path taken in the parameters plane. (b) Sequence of the billiard deformations $(X,Y)$ corresponding to the loop avoiding the diabolical point, as shown in (a). The red dot marks the point of accidental degeneracy, and the open dot marks the start and the end point of the loop.}
\label{Fig3}
\end{figure}

To visualise the Berry phase, we need to follow the transformation of the phase of the eigenstates as we traverse a contour in the plane of the two deformation parameters. Near-field imaging after a spectrometer allows measurement of the probability density of the eigenstates of a polariton billiard (see Methods). The Berry phase can then be determined without the need of interference experiment following the protocol from Refs. \cite{Lauber1994,Moon2009}. This protocol is more straightforward to implement in our experiment since both the eigenstates involved are two-lobed modes with a well-defined orientation of the nodal line. However, due to the finite energy linewidth, the deformation should be large enough to avoid the overlap between the two modes.

Figure \ref{Fig2} shows the transformation of the probability density and the phase of the two eigenstates when a loop enclosing the diabolical point is followed in the parameter plane. An arbitrary starting point in $X-Y$ plane is chosen and the initial phase is set by defining one lobe of the probability density as positive (zero phase) and the other as negative ($\pi/2$ phase). Each panel in Figure \ref{Fig2} therefore shows the  experimentally measured probability density distribution multiplied by the corresponding phase, with the $\pi/2$ phase step across the nodal line. Then we move to the nearest data point clockwise in the $X-Y$ plane and assign the relative phase in such a way as to ensure a smooth transition from the previous point. The process is repeated for the subsequent data points until a loop enclosing the diabolical point is completed. At the end of the loop around the diabolical point, the phase of each eigenstate has flipped, i.e. accumulated the $\pi$-phase shift, which corresponds to the geometric Berry phase. A second loop is therefore necessary to return to the original wavefunction.

\begin{figure}[ht]
\centering
\includegraphics[width=10cm]{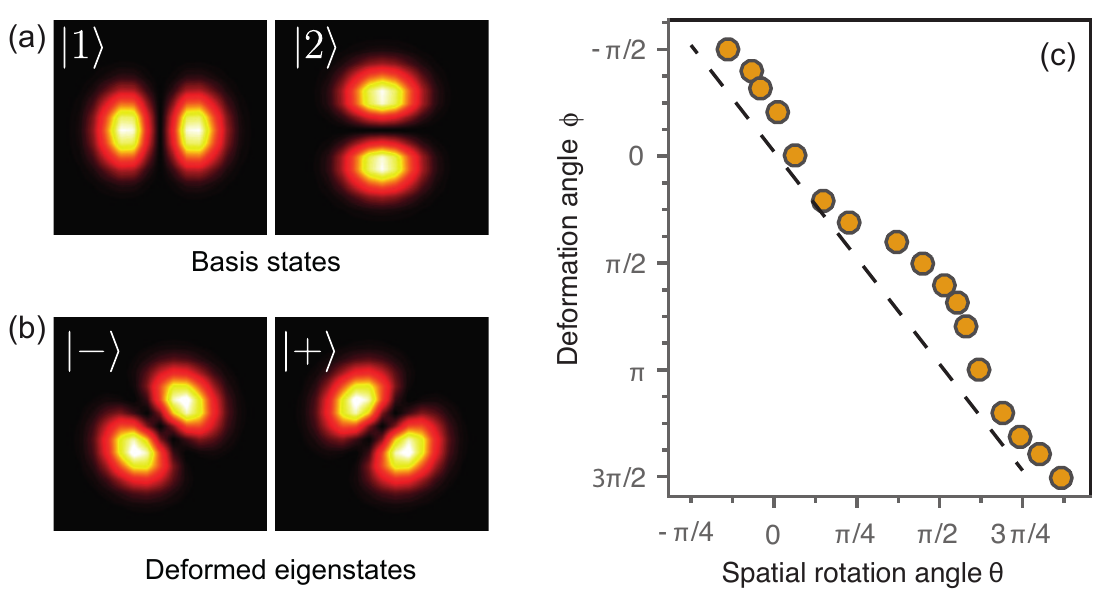}
\caption{Theoretically calculated probability densities (a) of the chosen basis states in Eq \ref{eq1} and (b) of the mixed states due to deformation angle $\phi=\pi/2$ showing rotation in real space by $\theta=\pi/4$. (c) Experimentally measured rotation angle $\theta$ of the deformed eigenstates as a function of the polar angle $\phi$ of the deformation parameters $X$, $Y$. Dashed line corresponds to the  relationship $\theta=\phi/2$ expected to arise due to an ideal diabolical cone shown in Figure \ref{Fig1}(a).}
\label{Fig4}
\end{figure}

A different loop in the plane of the two deformation parameters can be chosen to {\em avoid} the diabolical point, as shown in Figure \ref{Fig3}. Following the same procedure, one can see that there is no accumulated phase shift after completing a single loop, which is in contrast to Figure \ref{Fig2}. This suggests that there is indeed a true degeneracy in this two-level system, even though there are uncontrolled asymmetries in the billiard shape arising from imperfections of the optical pumping profile.

It is well known that perturbation of the billiard potential due to shape deformations causes mixing of the original eigenstates as shown in Eq. \ref{eq2}. Importantly, here this mixing is manifested as a rotation of the eigenstates in real space as observed from the evolution in probability density in Figures \ref{Fig2} and \ref{Fig3}. This rotation can be understood with the help of Eq. \ref{eq2}. Indeed, by choosing basis states $|1\rangle$ and $|2\rangle$ corresponding to the two-lobed modes in Figure \ref{Fig4}(a), the corresponding eigenstates of the deformed billiard $|+\rangle$ and $|-\rangle$ [Figure \ref{Fig4}(b)] can be viewed as the basis states with the nodal line rotated by the angle $\theta=\phi/2$, where $\phi$ is the polar angle in the plane $(X,Y)$ of the deformation parameters. For an ideal system modelled by Eq \ref{eq1}, the relationship between the two angles is linear [dashed line in Figure \ref{Fig4}(c)]. The experimentally measured relationship between the deformation angle and the rotation angle of the eigenstates shown in Figure \ref{Fig4}(c) displays a slightly nonlinear behaviour. This is primarily due to the elliptical cone shape of the energy sheets of the two states exploited in this experiment in contrast to the circular cone of the ideal model shown in Figure \ref{Fig1}(a). Nevertheless, it is this simple correspondence between the billiard deformation angle and the angle of rotation of the nodal lines for the two-lobe probability density distributions of the states $|1\rangle$ and $|2\rangle$ that enables us to observe the Berry phase directly, without the need for interferometry or numerical modelling \cite{Gao2015}. 



\section*{Discussion}

It should be noted that, due to the intrinsic open-dissipative nature of our system, the quantum billiards for exciton-polaritons are non-Hermitian \cite{Gao2015}. However, in contrast to Ref.~\cite{Gao2015}, here we do not change the thickness of the billiard walls [black area in Figure \ref{Fig1}(b), inset] and therefore do not actively manipulate the relative overlap of the different billiard modes with the gain region created by the optical pump. As a consequence, imaginary parts (linewidths) of the two energy eigenvalues used to construct the diabolical point {\em grow concurrently} and {\em nearly monotonically} as the parameters $(X,Y)$ are detuned from the point of exact degeneracy. This ensures that we are dealing with an equivalent of a Hermitian spectral degeneracy well modelled by Eq. \ref{eq1} rather than an exceptional point considered in Ref.~\cite{Gao2015}. This behaviour of the energy eignvalues and spectral linewidths in an exciton-polariton billiard is consistent with that observed in dielectric optical microcavities \cite{AnPRL2009,AnPRA2009}. Indeed, it was established \cite{AnPRA2009} that, in open resonators (classical wave billiards), the linewidth of two neighbouring levels increase monotonically as they move along the {\em diabatic lines}  [e.g., dashed lines on the diabolic cone in Figure \ref{Fig1}(a)] away from the point of degeneracy. Moreover, non-integrability due to geometric deformations leads both to simultaneous growth in coupling between energy levels (manifested in avoided crossings) and in decay rates (linewidth) \cite{AnPRL2009}. This is exactly the behaviour observed here as we move further away from the point of degeneracy (integrability). Therefore, in order for the simple theoretical model  Eq. \ref{eq1} to be applicable, we have to ensure that the values of the deformation parameters remain sufficiently small. Large deformation parameters result in larger deviation of the experimentally observed intersection of energy levels from the exact diabolical cone, which manifests itself in further departure from the linear dependence $\theta(\phi)$ (Figure \ref{Fig4}).

To summarise, we have visualised the quantum Berry phase by exploiting the existence of a diabolical spectral degeneracy in a deformed square billiard for macroscopic coherent waves of microcavity exciton-polaritons. By constructing a loop in the parameter space that encloses the diabolical point, we observed a $\pi$ geometric Berry phase shift of the eigenmodes. Consequently, no geometric phase was observed when the closed path avoided the diabolical point. We have also demonstrated that, with a proper choice of the basis, the two eigenstates in the non-separable system are a superposition of the degenerate states which depends on the deformation angle $\phi$ that rotates the original state by an angle of $\phi/2$. Our experimental observations agree well with a simple two-level Hamiltonian model near a diabolical point. 

Our work illustrates that, due to the opportunities presented by direct imaging of the condensate spectra and nodal structure of the spatial probability densities (wavefunctions) via the cavity photoluminescence, exciton-polariton billiards represent an ideal system for visualising and exploring geometric phase effects in quantum systems.


\section*{Methods}

In our experiment, exciton-polaritons are created in an AlAs/AlGaAs microcavity containing $12$ GaAs quantum wells ($\sim13$ nm wide each) sandwiched between distributed Bragg reflector mirrors (32/36 mirror pairs). The quasiparticles are excited by an off-resonant, linearly polarised pump beam derived from a continuous wave (CW) Ti:sapphire laser operating at $732$ nm, and driven to condensation above the threshold power of $\sim 0.079$ mW/$\mu$m$^2$ at $\sim 6$ K maintained inside a continuous flow microscopy cryostat. Details of the experimental apparatus can be found in Ref.\cite{Gao2015}

A quantum billiard for exciton-polariton waves is created by structural shaping of the optical pump using a digital micromirror device (DMD), as detailed in Ref. \cite{Gao2015} The pump shape reflected by the DMD mirror and re-imaged onto the surface of the sample through a high NA objective lens is shown in Figure \ref{Fig1}(b), inset. The inner length of the square billiard wall is $L=12$ $\mu$m, and the wall thickness is $4.5$ $\mu$m. Cavity photoluminescence resulting from decay of exciton-polaritons and release of coherent photons is collected via the same objective lens and analysed using a spectrometer and a camera. Analysis of the microcavity emission  therefore delivers the spectrum of exciton-polaritons [as seen in Fig. \ref{Fig1} (b,c)] and real-space images of the spatial probability distribution for the exciton-polariton condensate, as seen in Figures \ref{Fig1}-\ref{Fig3}. Above the threshold pump power, exciton-polaritons occupy multiple eigenstates of the optically defined trap  \cite{Tosi2012,Baumberg2013,Lagoudakis2013,Lagoudakis2015,Gao2015,Snoke2016} in the shape of the rectangular potential, which are resolved by analysing the spectrum of the photoluminescence with the aid of tomography. The errors in calculating the energy of the individual eigenstates arise from numerical fit of the spectral line and are therefore very small [see Figure \ref{Fig1}(c)].

The deformation of the billiard potential is achieved by reprogramming the DMD to reflect a different shape. In particular, the loops in the parameter plane displayed in Figures \ref{Fig2}(a,b) and \ref{Fig3}(a) are produced by the deformation sequences depicted in Figures \ref{Fig2}(c) and \ref{Fig3}(b), respectively.



\section*{Acknowledgements}

This work is supported by the Australian Research Council and the State of Bavaria. Discussions with T.C.H. Liew and K.Y. Bliokh are gratefully acknowledged.

\section*{Author contributions statement}

E.E. and E.A.O. conceived the experiment,  E.E., T.G., and  A.G.T. conducted the experiment, S.B., M.K., C.S., and S.H. fabricated the microcavity sample, E.E., T.G., A.G.T. and E.A.O. analysed the results, E.E. and E.A.O. wrote the manuscript.  All authors reviewed the manuscript. 

\section*{Additional information}

The authors declare no competing financial interests.

\end{document}